\documentstyle[12pt]{article}
\begin{document}
\begin{center}
{\large\bf The source of irreversibility in macroscopic dynamics}\\
\vskip 5mm
X de Hemptinne
\vskip 5mm
\end{center}

\section{Introduction}

Distortion of macroscopic systems initiates spontaneous and
irreversible dynamic processes tending to restore the previous state
of equilibrium or to establish a new one. There is no doubt that in
this context the word irreversibility points to the fact that the
initial perturbation is not regenerated spontaneously. Unless promoted
by some external action, it belongs to the past. Concerning the
definition of an equilibrium state, the scientific literature is less
specific, although everyone has a personal feeling about what this
word should mean. The most common definition, but not the only one,
relates equilibrium conditions to the extremum of certain
thermodynamic potentials [1]. It will be shown that this definition is
insufficient.

Historically, the debate on irreversibility started with Joule's
experiment, when he allowed an ideal gas to expand spontaneously from
a small vessel into a larger one. The process was carried on inside a
calorimeter. Expansion proceeds apparently without exchange of heat
with the outside world. If by {\it isolation} it is meant that the
system exchanges nor heat nor matter with the outside world, following
the tradition in conventional thermodynamics, it must be concluded
that spontaneous and irreversible expansion of the gas is the
expression of a fundamental property of isolated systems. In the
literature, the words {\it closed} and {\it open} system are sometimes
used to indicate whether they are isolated or not. The latter
expressions are however ambiguous and they will not be used here.

In passing, let it be stressed that the meaning the dictionary gives
to the word {\it isolation} is stronger than that suggested above. It
points to objects that are left alone and excludes therefore all
possible interactions whatever with other systems. By contrast, the
weak thermodynamic meaning allows elastic collisions with the
boundaries (energy conservation).

The word {\it irreversibility} too has different connotations. At
first, taking the colloquial meaning suggested above (no
spontaneous recurrence of previous conditions), the word does not
specify whether or not the memory of earlier perturbations remains in
the system under some hidden form.

Secondly, in the context of mechanics, the word expresses the response
of the dynamics to the fictitious mathematical operation consisting in
the sign reversal of the variable {\it time}. With this definition,
systems for which the equations of the motion are not invariant with
respect to changing $t$ into $-\,t$ are irreversible.

Closely connected to the latter and referred to mainly in numerical
simulation of relaxing systems, irreversibility concerns also the
effect of artificially reversing the sign of the time increment ${\rm
d} t$, starting at some given instant in the course of the dynamic
process. If this mathematical manipulation does not reproduce the
initial conditions as an echo, the dynamics is called irreversible.
The variety of definitions of this important keyword leads to frequent
confusions.

In trying to rationalize irreversible dynamics of isolated systems,
Ludwig Boltzmann suggested that the directionality with respect to
time would be caused by inter-particle collisions or interactions.
Considering that for his contemporaries {\it isolation} was understood
almost {\it stricto sensu} (only elastic reflections at the boundaries
allowed), thereby validating Hamiltonian mechanics, and that the
mathematical definition of {\it irreversibility} (time-symmetry) was
withhold, Boltzmann's suggestion unlocked a storm of dispute.
Contrasting with the initial turmoil, a majority of authors seems now
to be confident in Boltzmann's kinetic theory [2,3,4], although
intellectual discomfort remains apparent [5,6]. The question is still
active in the scientific literature.

The microscopic deterministic laws that govern collisions are strictly
symmetrical with respect to sign reversal of the variable $t$. For
justifying the apparently unexpected violation of the time-symmetry of
global dynamics of many-particles systems, it is assumed by many that
this is the consequence of the extremely large number individual
interactions and also of their complexity. Progress booked in recent
years both in mathematics and in statistics is alleged to support the
suggestion. Chaotic motion is also are frequently referred to in view
of strengthening the demonstration. However, arguments imported from
neighbouring domains of sciences without double check concerning
conformity of the definitions and compatibility of the initial
assumptions lead to ambiguous conclusions.

Symmetry ranks undeniably among the most robust general features of
nature, resisting all mathematical manipulations, limit operations or
transitions from discontinuous to continuous models. This fact is
confirmed in all branches of sciences, like spectroscopy,
crystallography, geometry etc. Considering the extreme strength of
symmetry properties, mathematical manipulations consisting merely in
limit operations performed on scaling factors in order to explain
breaking of the time-symmetry in the global dynamics of isolated
systems in the many-particle limit cannot be justified.

This paper is to show that the source of irreversibility is to be
fetched in the unavoidable interaction with the outside world. To that
end the discussion is structured as follows. In the first section, the
traditional Joule experiment is analyzed. It stresses that the global
process consists in the superposition of two fundamentally different
independent mechanisms: one is conservative and the other one
dissipative. Dynamic properties of the conservative part and of the
dissipation part of many- particles mechanics are the subject of the
two sections to follow. Finally, the entropy and its change in non-
equilibrium processes are examined.

\section{Joule experiment}

In the literature, discussion concerning the directionality of the
time's arrow is often introduced intuitively on the basis of a
simplified description of Joule's experiment. We consider an isolated
box consisting of two compartments, the parts being filled with gas at
different pressures. The observed long time evolution towards
homogeneous distribution following the rupture of the division is
taken as picturing the irreversible behaviour of the global dynamics.
In this context irreversibility means non-recurrence of the initial
conditions.

Joule's purpose was to investigate possible heat exchange with an
external calorimeter as the result of spontaneous expansion. With an
ideal gas, if no mechanical work is allowed to be performed, when the
system has reached its final state, he observed no net heat exchange
with the surroundings. His observation did however not preclude
possible fluctuating exchange with zero balance. For the observer, the
system behaves apparently as if it was isolated according to the usual
thermodynamic meaning of the word.

Let us make the experiment more realistic by examining instead the
effect of puncturing an air inflated bladder inside either an acoustic
reverberation hall or an anechoic chamber. The conditions may be
approximated simply by comparing the properties of a completely
furnished room and that of the same room stripped of its furniture,
curtains and rags. In the two cases the excess air contained in the
bladder disseminates spontaneously throughout the rooms but the
subsequent process is very different indeed. In the reverberation hall
an acoustic perturbation is created and, the better the walls'
reflecting quality, the longer it remains. By contrast, in the
anechoic room, the perturbation vanishes promptly.

The first question the double experiment raises is that of the
definition of the state of equilibrium. The final state reached in an
ideal reverberation room allows some energy to be stored in a
coherent collective motion (acoustic perturbation) where it remains as
the memory of the initial conditions. With walls shaped in
particularly favourable forms, the initial information may even be
partially retrieved as echos. By contrast, in the anechoic room,
memory of the past is soon forgotten. Clearly only the latter
condition may be considered as an equilibrium state.

The air in the rooms is the same, and therefore also the frequency and
the quality of the inter-particle collisions. The only difference
between the two experiments is the nature of the walls. If we define
relaxation as being the process whereby the distorted system reaches
equilibrium, the role of the interaction with the neighbourhood
(walls) is trivial. If equilibrium is to be reached, the system cannot
be strictly isolated.

Let us examine the sequel of events leading to final relaxation.  When
the membrane is ruptured, a stream of gas is ejected from the vessel
at the higher pressure, thereby forming a collective motion of the
particles. Energy that is transferred into the jet is subtracted from
the initial thermal supply (adiabatic expansion, constant entropy). As
a result, the system's temperature drops \footnote{not the average
kinetic energy but $(\partial S/ \partial E)^{-1}$}.

The wall opposite the puncture reflects the jet and turns the
collective motion progressively into an acoustic perturbation with the
same energy. The spectrum and phases of this motion reflect the
initial conditions and the shape of the reverberating walls
(coherence). While dissemination of the particles throughout the
system occurring during this part of the motion is irreversible
according to the {\it no recurrence} definition (known in mathematics
as mixing [7]), information about the past is not forgotten, no matter
how intricate the motion of the particles may be.

Final relaxation of the coherent motion starts now. It consists in
thermalizing the energy stored in the acoustic perturbation, thereby
balancing or neutralizing the initial temperature drop. When full
equilibrium has been reached, the information about the initial
conditions is completely lost.

\section{Conservative trajectories}

The scenario discussed above shows that the global motion of
macroscopic systems is governed partly by Hamiltonian dynamics (jet
forming and dissemination or mixing) and partly by dissipation
interaction with the neighbourhood (final relaxation). We focus now on
the Hamiltonian part, neglecting whatever would make strict isolation
ineffective. If the Hamiltonian is not an explicit function of time,
the dynamics it supports is by definition conservative and
deterministic \footnote{Conservative Hamiltonian mechanics is
deterministic but friction and other irreversible macroscopic
transport properties are too, without being conservative.}.

A system is said to be conservative if the force field is such that
work done around a closed orbit is zero [8]. Physically it is clear
that dynamics cannot be said to be conservative if friction or other
dissipation forces are present.

In classical mechanics, determinism or causality is the property
according to which, if two dynamical systems have the same laws of
motion and are in the same dynamic state at some particular time
$t_0$, then they must be in the same dynamic state at all times [9].
The corollary is that no more than one trajectory passes through each
point in phase space, or else that different phase space trajectories
never cross. Let it be stressed that this discussion uses the word
{\it trajectory} as a reference to global many-particles motions but
never to the individual paths of separate particles.

In the context just mentioned, {\it dynamic state} is the same as the
word {\it microstate} used by some authors. It represents a point
${p_1...p_{3N} , q_1...q_{3N}}$ in the many-particle phase space. A
phase space trajectory is the succession of points (or phases) that
pictures the system as time goes on. Hamilton's canonical equations
are its parametric equations. Considering that there is only one
trajectory passing through a given phase point, there is a unique
relationship between {\it microstate} and {\it trajectory}. Unless the
investigation concerns singular properties of individual phase points
along a given trajectory, the definition of {\it microstate} may
therefore equally be covered by the complete trajectory. The two words
represent the same reality.

Action referred to as {\it mixing} is often quoted in the literature
[7]. It describes the way our perception of the initial conditions
changes as time goes on assuming a deterministic and conservative
environment. This perception concerns the distribution of individual
phase points along a single trajectory in phase space. In the above
mentioned example of an expanding gas, dissemination of the particles
throughout the volume is synonymous to mixing.

Let it be stressed that, parallelling ambiguity concerning the
definition of true equilibrium, there is some confusion in the recent
literature concerning the alleged relaxing property of mixing. The
example of the expanding gas suggests that, contrasting with mixing,
relaxation belongs to the second step of the global process.

Mixing does apparently not require inter-particle collisions or
interactions [7]. The relevant dynamics may be exemplified by a system
of many non-interacting particles, translating back and forth along a
line between two boundaries (distance $2D$), where reflection of the
individual motions is elastic (energy conservation).

We consider a many-particles system and assume that the velocity
distribution is Gaussian (Boltzmannian). Hence
\begin{equation}                            
g(v) \approx \exp (- \beta \, {{m v^2}\over {2}}).
\end{equation}
The relevant parameter $ \beta$ is not the reciprocal
temperature \footnote{In this model the system is forced to remain on
a single global trajectory. As it will be shown later, its entropy is
therefore zero and the concept {\it temperature} $(\partial S/
\partial E)^{-1}$ is meaningless.}.

We assume that the initial conditions determining the single global
trajectory in phase space are represented by a giant density
fluctuation at some given position along the line. It would be
represented by a $ \delta$-function. It is then easy to compute the
density distribution at any later time. The sharp fluctuation
disappears, leading at long times to a nearly flat particle
distribution, as if the initial perturbation had relaxed. However,
tiny irregularities persist indicating that the initial fluctuation
has turned into noise.

If the number of particles is high, the initial fluctuation does not
recur spontaneously after a reasonable delay (Poincar\'e recurrence).
The process responds to the colloquial meaning of irreversibility.
However, according to the mathematical or mechanical definition of the
word, the relevant dissemination is by no means irreversible. Indeed,
if the sign of the velocities of all the particles is artificially
reversed at any instant, the global trajectory defined by the initial
fluctuation is made to run in the opposite direction and the latter is
reproduced as an echo. The information represented by the initial
conditions was still present in the system although, due to the
diversity of the individual velocities, the sharp starting impression
has been progressively hidden to the observer.

The impression the observer has about the system's conditions and its
change in the course of time may be expressed by the evolution of the
position of the centre of mass. Let $X(t)$ be this position. If
$x(v,t)$ is the position of any particle with velocity $v$ at time
$t$, $X(t)$ is defined as
\begin{equation}                         
{{X(t)}\over {X(0)}} = {{\int g(v)\,x(v,t)\,{\rm d}v}\over{\int
g(v)\,x(v,0)\,{\rm d}v}}.
\end{equation}
Starting from the initial giant fluctuation, supposing this is
eccentric with respect to the boundaries, as time goes on, the
particles disseminate and the centre of mass moves towards the
system's geometrical centre $(X(t=\infty)=0)$. By computation, the
general expression for the dynamical evolution of $X(t)/X(0)$ is
easily shown to be
\begin{equation}                  
X(t)=X(0) \exp [- \phi (t^2)].
\end{equation}
There are no odd powers of $t$ in the dynamics. If the initial
fluctuation is situated at $x=\pm 0.8 D$, the result is almost a pure
Gaussian:
\begin{equation}
X(t)-X(\infty) \approx X(0) \exp [- 0.62 \left(t \over \tau \right)^2],\\
\end{equation}                       
$$\tau={\sqrt{{\beta m}\over 2}}D.$$
For other initial conditions the decay is not very different. In all
cases, the very fast Gaussian-like decay is by no means comparable to
the exponential or multi-exponential dynamics of conventional
relaxation.

The general conclusion is not modified if collisions or chaos
generating elements (e.g. Sinai billiards) are included in the
dynamics. Only computation becomes harder. Possible rounding-off
errors produced at every iteration step cause an artificial relaxation
that has nothing in common with exact dynamics.

\section{Dissipation and fluctuations}

Conservative motion defines trajectories in the many-particle phase
space. By forbidding transitions between different trajectories it
preserves the memory of the initial conditions. For transitions
between trajectories to occur, the conservative motion must be
perturbed. This occurs every times the system interacts with its
boundaries or with whatever represents its neighbourhood, like the
ubiquitous thermal electromagnetic radiation and gravitation fields.
Export and import of information being uncorrelated, the transitions
cause loss of information and irreversible relaxation of the initial
single microstate.

A {\it macrostate} is by definition the observational condition of a
system where some or perhaps all the trajectories (microstates)
belonging to a narrow band of energy hypersurfaces or energy shells in
phase space are accessible.

With systems of translating particles, interaction with the outside
world occurs every time a particle hits the walls. Every collisions of
any particle with a wall interrupts the running canonic global
trajectory and starts a new one with possibly modified initial
conditions. The average lifetime of the trajectories depends on the
impact frequency of particles with the walls.

More assumptions are needed to predict the effect of collisions with
the walls. If the latter are perfectly rigid, so that they behave as
virtual particles with infinite masses, only momentum is transferred.
When the impact is over, the wall has gained momentum from the
particle but no velocity, because its mass is infinite. In the same
time the wall has given to the particle an equal amount of momentum in
the opposite direction, allowing the new trajectory to start in
conditions that are rigourously correlated with the previous one. The
conservative character of the motion remains.

Real atoms and molecules in the walls have finite (effective) masses
and they oscillate about their equilibrium positions. The result is
that, contrasting with the picture drafted above, transfer of momentum
at every impact is associated with some transfer of energy. Given the
initial trajectory of the incident particle, the exact return path is
unpredictable, because the motions of the collision partners are
uncorrelated. Some impacts with the wall atoms hit the incident
particle harder and some less.

If the wall atoms oscillate about fixed positions, the average
transfer of energy over a period of time that is long when compared to
the collision periodicity with the walls is nil, while every single
impact imports fluctuations about this average value.

The average lifetime of conservative trajectories in phase space is an
important dynamic parameter. Its value equals the reciprocal of the
average global periodicity of the collisions with the walls times an
efficiency parameter. Referring to Joule's experiment mentioned above,
with hard reflecting walls the efficiency parameter is extremely
small; with soft walls it approaches 1.

With experiments where the resolution time is made exceptionally fine
(e.g. very short laser pulse experiments), the parameters of
individual running motions are accessible and well defined single
trajectories may be observed. The particular phase or the location of
the system's representative point along a defined trajectory in phase
space is then observable. Such systems are practically isolated during
the time of observation. Measurements on systems where immediate phase
information remains pertinent follow the traditional laws of
conservative mechanics. Statistics and thermodynamics are meaningless
to them.

If by contrast the experimental resolution time is long enough, as it
is the case with most relaxing properties, the system is allowed to
perform stochastic jumps between the accessible trajectories by
exchange of momentum, energy etc. with its surroundings. Phase
information along single trajectories or even the definition of
particular trajectories in phase space become irrelevant. Measurements
are then restricted to statistics. Description of the system and
eventually of its dynamics implies the use of thermodynamic arguments.

\section{The Entropy}

The macrostate of thermodynamic systems is defined unambiguously if
the complete set of parameters representing the constraints, either
external or internal, implied by the system's particular observational
state is mentioned both qualitatively and quantitatively . Any
function determined completely by such parameters is a function of
state.

In 1865 Clausius discovered a function of state that changes when heat
(energy, excluding work) is exchanged reversibly with the environment.
This function, Clausius' entropy, is defined as a differential.
Assuming a reversible process, the definition reads
\begin{equation}                        
{\rm d} S \geq {{{\rm d} Q}\over T},
\end{equation}
where $T$ is the system's temperature. Equal sign refers to reversible
processes. This fundamental experimental definition implies net
transfer of heat and therefore interaction of the system with its
environment.

Clausius' requirement that the process would be reversible for the
equal sign to be valid means that the system may not depart from
equilibrium during the whole process. This specification is rather
ambiguous, as it relates to a property (equilibrium) that is by itself
insufficiently defined. Instead, let us suggest that the word
reversibility used in this context implies that no coherent or
collective motion is allowed to be generated by the process or that it
has been made to relax.

The physical meaning of the state function entropy and especially the
discovery that it increases when the system is the subject of
spontaneous or irreversible processes in apparently isolated
conditions (conservation of energy and matter) has intrigued many
physicists and philosophers. Having established a relationship between
the entropy and some kind of observational probability, some try to
attribute to this concept an anthropomorphic character [10]. It would
be the measure of our personal lack of information concerning the
system's conditions. This strange suggestion that personalizes a
function of state indicates that profound confusion prevails
concerning the definitions.

In statistical mechanics, which is the theoretical branch of
thermodynamics, the definition of entropy goes back to Ludwig
Boltzmann. It is summarized by his famous equation
\begin{equation}                               
S=k_B \ln [W(A)].
\end{equation}
For the inventor, $W(A)$ meant {\it wahrscheinlichkeit} which is
probability. Digging for the realities hidden behind this word may
lead to some controversies but, using the same initial letter, most
authors wisely prefer now the English {\it weight of the given
observational state}. The latter is interpreted as the volume
accessible to the motion in phase space, given the set of constraints
(represented here by the collective variable $A$) that describe the
system's particular macrostate (observational state). Let it be noted
that an equilibrium macrostate is usually defined by its total energy
$E$, particle number of any sort $N_{\rm r}$ and physical volume $V$,
which are the traditional microcanonical variables.

As such, the definition of $W(A)$ misses normalization. This leads to
the introduction of an arbitrary constant in the entropy. Planck
filled the blank [11] by suggesting that $W(A)$ is the total number of
independent quantum states (quantum state = microstate) compatible
with the given macrostate. The latter represents the third law of
thermodynamics according to which, if only one state is accessible,
the entropy is zero.

Rephrased with Planck's conclusion, Boltzmann's entropy clearly
favours a quantum description of the motion. Unfortunately, quantum
mechanics does not refer directly to phase points or trajectories in
phase space as does classical mechanics. By contrast, the quantum-
classic correspondence implies that every global quantum state is
pictured in the classical phase space by a finite region with a $6N$
dimensional phase volume measuring $h^{3N}$. The number of available
classical trajectories respecting the state defining constraints
equals the ratio of the accessible phase space volume to $h^{3N}$.

For Boltzmann's entropy to be a pertinent function of state,
accessibility of several (many) quantum states or trajectories is
required. In a strictly conservative environment, the dynamics being
described by a single global trajectory, no matter how intricate
(chaotic) this may be, the necessary transitions between different
trajectories or quantum states are not allowed. Then, according to the
definition, the entropy is zero and it does never change. This
conclusion is consistent with Liouville's theorem that claims
conservation of the measure in phase space when the mechanics is
conservative.

Relaxation implies relief of constraints. It opens the way to an
enhanced choice of quantum states or trajectories. Accessibility of
more trajectories increases Boltzmann's entropy.

Accessibility implies swift transitions between many trajectories or
quantum states during the observation period. This depends on fast
uncorrelated action of the environment with exchange of mechanical
properties (momentum, energy). As a corollary and as expected by the
statistical nature of the thermodynamic functions it appears that the
definition of the entropy implies averaging over the time. The time
resolution linked to the definition of the entropy is the average life
time of the conservative trajectories.

In view of the high impact rate of macroscopic systems with their
boundaries, the period of time over which averaging is required may be
extremely short indeed. For complex relaxing systems, equilibration of
the different constraints with the outside world may not be equally
fast. The fastest relaxing process concerns usually thermalization of
the translational energy. The relevant intensity (translational
temperature) reaches its environment value. The residual dynamics is
governed by the least efficiently exchangeable properties.

In describing equilibrium states (e.g. for a one-component gas), the
extensive variables mentioned traditionally are $E$, $V$ and $N$. That
are the basic microcanonical constraints. In order to specify
unambiguously non-equilibrium macrostates, where additional
constraints prevail, additional extensive properties must be included.
This may be for example the momentum associated with a possible
collective or coherent motion of the system, in which some of the
total energy is stored (e.g. the jet or the acoustic motion in Joule's
experiment). Many other possible distortions with respect to
equilibrium may be envisaged, like moments of the energy or density
distribution, etc.

Let the list of the extensive properties of a macroscopic system
defining a particular macrostate be written in general $X_{\rm r}$. By
differentiating the entropy with respect to the set of $X_{\rm r}$, we
get by definition the set of conjugate intensive variables or
intensities $\xi _{\rm r}$.
\begin{equation}                        
{\rm d} S= \sum _{\rm r}\, {{\partial S}\over{\partial X_{\rm
r}}}\,{\rm d} X_{\rm r}=-k_B\, \sum_{\rm r} \,\xi_{\rm r}\,{\rm d}
X_{\rm r}.
\end{equation}
The temperature $(\partial S/\partial E)^{-1}$ and the chemical
potential $-T(\partial S/\partial N)$  are not new. In non-equilibrium
conditions, the equation generalizes the definition, covering now also
the intensities conjugate to the additional non-equilibrium
constraints representing the particular macrostate. The non-
conventional procedure consisting in mentioning the latter intensities
has been introduced elsewhere [12].

Equation (7) is Gibbs' celebrated differential equation, generalized
by including the non-equilibrium constraints explicitly. In a
simplified model of the spontaneously expanding jet referred to above
(velocity of the collective motion: ${\vec v}$), the new version of
Gibbs' equation reads
\begin{equation}                         
{\rm d} S={{{\rm d} E}\over T}+{p\over T}{\rm d} V-{\mu \over T}{\rm
d} N-k_B\,{\vec \sigma}\cdot{\rm d} {\vec{\cal P}},
\end{equation}
where ${\vec{\cal P}}=Nm{\vec v}$ is the collective momentum and
${\vec\sigma}$ the conjugate intensity. It may be shown (see reference
[12]) that ${\vec\sigma}={\vec v}/k_BT$. In the last term, change of
the collective or coherent energy is easily recognized. We have
therefore equivalently
\begin{equation}                         
{\rm d} S={{{\rm d} E}\over T}+{p\over T}{\rm d} V-{\mu \over T}{\rm
d} N-{1\over T}{\rm d} (coherent\; energy).
\end{equation}

Conservation of energy throughout the expansion process makes ${\rm d}
E=0$. During the adiabatic dissemination or mixing period, the second
term (work performed by the expansion) is very exactly balanced by the
last contribution (energy stored in the coherent motion), making ${\rm
d} S=0$. This result is in full agreement with Liouville's theorem in
isolated (conservative) conditions. Final relaxation involves
transformation of the coherent motion into thermal energy. When this
has been achieved, thanks to momentum rephasing of the individual
particles on inelastic exchange at every impact with the boundaries,
the last term vanishes and integration of the Gibbs equation yields
the correct final equilibrium entropy after expansion.

The present discussion concerning the two distinct mechanisms that are
involved in non-equilibrium processes resolves endless debates on the
relative merits of Boltzmann's and Gibbs' approaches to the total
entropy change [11]. Their conclusions refer indeed to different steps
of the process. Gibbs considers the adiabatic one and concludes
correctly that ${\rm d} S=0$ on mixing. By contrast, Boltzmann's
entropy change relates to the relaxation step but he assumes
incorrectly compatibility of this step with isolation of the relaxing
system.

If the contribution relating to the non-equilibrium constraint is
omitted in Gibbs' differential expression, one is forced to replace
the equality sign in the equations by the $\geq$ sign, in agreement
with Clausius' original inequality.

\section{Conclusions}

The main conclusion to be drawn from the discussion above is that
dissipation of the non-equilibrium constraints of macroscopic systems
involves interaction with the outside world.  The word {\it
irreversibility} is ambiguous. If it is understood in the sense of
non-recurrent change in the course of time, conservative {\it mixing}
belongs indeed to that class of phenomena. This sort of
irreversibility might however be qualified as apparent or weak. Its
end- point is indeed not thermodynamic equilibrium. Memory of the
initial conditions is still present under hidden form. By contrast,
the opening of the system to a broad choice of new accessible
conditions by exchange with the environment, thereby replacing and
perhaps even neutralizing the former ones ensures the strict or strong
irreversible character of dissipation.

An objection sometimes raised against the privileged role of the
neighbourhood is that the proposal merely moves the difficulty step-
wise further, while it is often assumed that the universe itself
should be isolated. To this it must be answered that extrapolating the
conclusions valid at our observational level to the whole universe is
mixing up physics and metaphysics. The two are respectful domains of
sciences but their tools and objectives are very different indeed. The
first goal remains focusing on facts that are directly accessible to
the experiment rather than elaborating a theory that is beyond our
reach.

The properties of the environment are crucial, also in defining the
state of equilibrium. If the system's container is moving, equilibrium
conditions imply that the system would be moving too. At equilibrium,
collective properties of the system and of the surroundings are
related. The intensities (differentials of the entropy with respect to
the values of the exchangeable extensive properties) in the system and
in its neighbourhood are equal. It is not correct to define the
equilibrium state on the only basis of the extremum conditions of
functions of state.

The objective of Boltzmann's equation
\begin{equation}                         
{{\partial f_1}\over{\partial t}}=-{p\over m}{{\partial f_1}\over
{\partial q}}+ {\cal C}(f_1,f'_1)
\end{equation}
was to rationalize the relaxation processes observed in macroscopic
systems assumed to be isolated from the outside world. It was obtained
from intuitive arguments with the purpose to justify irreversible
dynamics. It describes the conservative flow of a density of points in
a single particle phase space, perturbed by an interaction term. The
latter represents the modification to single-particle trajectories
brought about by inter-particle collisions. The theory of Sinai
billiards shows indeed that collisions do perturb individual paths,
making them even strongly cahotic, but the global motion remains
strictly symmetrical with respect to the sign reversal of $t$. This
contrasts with Boltzmann's perturbation term. Boltzmann's procedure is
therefore strongly conflicting with the first principles of mechanics.

Contrasting with mixing, dissipation to the surroundings is an
exponential or possibly a multi-exponential process. Let us assume
that the system's particular non-equilibrium state is related to an
extensive constraint represented by a given collective motion. This
represents part of the total energy. By definition, this component of
the energy is shared proportionally by the individual particles. At
every individual impact of a particle with the boundaries, the part of
the collective energy that is carried by the relevant particle (or a
proportional part of it) is thermalized. The rate of change of the
non-equilibrium constraint is therefore proportional to its actual
value in the system, leading to the expected exponential decay. This
simplified model is worked out completely elsewhere [12] and
generalized to the transport coefficients (viscosity, heat conduction
etc.) and other relaxation phenomena.

Transport coefficients belong clearly to the realm of dissipative
dynamics. They are meaningless in isolated systems. It must therefore
be stressed that the many types of phenomenological equations that
describe non- equilibrium dynamics and where such coefficients are
introduced, often on intuitive arguments (Navier-Stokes equations,
Fokker-Planck equations etc.) all relate to non-isolated systems.


\end{document}